\documentclass{article}

\usepackage{amsbsy,amssymb,amscd,amsfonts,amstext,amsmath}
\usepackage{delarray,graphicx,slashed, wrapfig, fancybox} 

\usepackage[utf8]{inputenc} 
\usepackage[T1]{fontenc}    
\usepackage{hyperref}       
\usepackage{url}            
\usepackage{booktabs}       
\usepackage{newtxtext,newtxmath}
\usepackage{nicefrac}       
\usepackage{microtype}      
\usepackage{enumerate}
\usepackage{color}

\newtheorem{theorem}{Theorem}[subsection]
\newtheorem{proposition}[theorem]{Proposition}
\newtheorem{lemma}[theorem]{Lemma}
\newtheorem{corollary}[theorem]{Corollary}
\newtheorem{remark}[theorem]{Remark}

\title{Algorithmic Complexities in Backpropagation and Tropical Neural Networks}

\author{
  \"Ozg\"ur Ceyhan   
  \vspace{.3cm} \\
  \texttt{ozgurceyhan@gmail.com}}

\begin{document}

\maketitle
 
\begin{abstract}
  In this note, we propose a novel technique to reduce the algorithmic complexity of neural network training by using matrices of tropical real numbers instead of matrices of real numbers.   Since the tropical arithmetics  replaces multiplication  with addition, and addition  with   $\max$, we theoretically  achieve several order of magnitude better constant factors in time complexities in the training phase.  

The fact that we  replace the field of real numbers with the tropical semiring of real numbers and yet achieve the same classification results via neural networks come from deep results in topology and analysis, which we verify in our note.  We then explore artificial neural networks in terms of tropical arithmetics and tropical algebraic geometry, and introduce  the multi-layered tropical neural networks as universal approximators.  After giving a tropical  reformulation of the backpropagation algorithm, we verify the algorithmic complexity is substantially  lower than the usual  backpropagation  as the tropical arithmetic is free of the complexity of usual multiplication. 

\end{abstract}

\section{Introduction}

The main strategies for building capable and more efficient neural networks can be summarised as
\begin{enumerate}[(I)]
 \item Developing and manufacturing more capable hardware; 
 \item Designing smaller and more robust versions of neural networks that realise the same tasks;
 \item Reducing the computational complexities of learning algorithms without changing the structures of
   neural networks  or   hardware.
\end{enumerate}
The approach (I) is an industrial design and manufacturing challenge. The approach (II) is
essentially the subject of network pruning. This note will    focus
only on a theoretical approach on (III) based on tropical arithmetics and geometry.

The main training method of multi-layered neural networks is the backpropagation algorithm and its
variations (see, for instance  \cite{ruder} for variants of  backpropagation    and their properties).  The backpropagation is essentially a recursive gradient 
descent technique that works on large matrices.  However,  the sizes of the matrices determined 
by the initial parameters of neural networks are enormous: The state-of-the-art applications may
have $10^6$-$10^7$ neurons, therefore, have as many  parameters as  the pairings of these
neurons,  see \cite[\S 1.2.4]{DBBNG}. Many neural network implementations require adequately large
computational resources when the scale of computations is that big, see \cite{DBBNG} and references
therein.  Naturally, since the computational resources required are large, any reduction in the
algorithmic complexity of elementary operations is going to provide substantial advantages.  To
this end, we would like to eliminate the basic arithmetic operation of multiplication, which is
algorithmically more complex than addition, subtraction and $min/max$, without changing the nature
of the classification problem that we wish to realise with a neural network.

We propose to reach our goal by using  tropical arithmetics and tropical geometry.  While
tropicalisation is a relatively new concept, its core idea has been used in engineering for decades
in areas such as the linear control theory and the combinatorial optimisation.  The effective use
of tropicalization in mathematics goes back to Viro in the 80's where he constructs real algebraic
varieties with prescribed topology to address Hilbert's 16th and related problems~\cite{viro}.
Further studies on Viro's method revealed the tropical semiring, an algebraic structure whose
arithmetics is devoid of multiplication, is the key algebraic structure behind Viro's
results~\cite{viro1}.  In this note we claim that the tropical geometry also provides a suitable
setup to construct   the backpropagation algorithm with substantially better complexity.

After briefly describing the sources of the algorithmic complexities in backpropagation techniques
in \S 2, and introducing the basic notions in tropical arithmetics and tropical geometry in \S 3.1,
we first define the tropical limit of the rectified linear unit (ReLU) in \S 3.2, and explore the
multilayered feedforward neural networks using this tropical unit in \S3.3.  We show that these
tropical neural networks have the properties of the universal approximator as the deep ReLU neural
networks. The last section \S 4 focuses on the topological realisation of the classification
problems via backpropagation based on the tropical semiring. In \S 4.1, we first reformulate the
classification problem as a topological problem of exploring the connected components of the
complement of the zero locus of a function approximated by a neural network. Finally, we introduce
the tropicalization of the backpropagation algorithm in \S 4.2, that essentially approximates the
original classification problem.  We conclude that, while we achieve an algorithm which
approximates the original problem with less algorithmic complexity, it does not require substantial
recoding, only the replacement of linear algebra packages with their tropical versions.

\subsection{Short history of this note and its current state}

This note essentially summarises the results presented by the author at {\it S\'eminaire "Fables G\'eom\'etriques"}
in University of Geneva on December 9, 2016.\footnote{  \texttt{unige.ch/math/tggroup/doku.php?id=fables} }. The version
at hand has been submitted for publication in 2017.  Since the completion of this note, tropical neural networks
started to attract considerable attention and  there has been a number of publications appeared. For instance, 
 {\it Tropical geometry of neural networks} appeared in ICML 2018 by Zhang et al was not available to the author when this 
 paper is written. This note is kept as it was initially submitted and no new results has been incorporated.


\section{Algorithmic complexity of backpropagation}

\subsection{Backpropagation in a nutshell}

Assume we have a multi-layered neural network with the weight matrix
$\mathbf{W}^{(k)} = \left[ w^{(k)}_{ij} \right]$ connecting the layers $k-1$ and $k$.  The
backpropagation algorithm aims at minimising a predetermined error function $E$ by using an
iterative process of gradient descent.  In this algorithm, one calculates the gradient matrix
$$ 
(\nabla E)^{(k)} = \left[ \frac{\partial E}{\partial w^{(k)}_{ij}} \right], k = 0,\dots, l
$$ 
for
training data, and adjusts the weight matrix $\mathbf{W}^{(k)}$ by adding a correction term
$\Delta \mathbf{W}^{(k)} = - \epsilon \nabla^{(k)} E$ in order to minimise the error function
iteratively.  The backpropagated error on the $k$-th layer is computed via an iterative matrix
multiplication
\begin{equation} \label{eqn_error}
\Delta  \mathbf{W}^{(k)}  = \mathbf{D}^{(k)}  \mathbf{W}^{(k+1)} \cdots  \mathbf{D}^{(l)}  \mathbf{W}^{(l+1)}  \mathbf{e}
\end{equation} 
where for each layer $k$, the matrix $\mathbf{D}^{(k)}$ is the diagonal matrix composed of the
derivatives of the activation function with respect to its arguments, and the vector $ \mathbf{e}$
contains the derivatives of the output errors with respect to the arguments of the last layer.  For
details, see for instance \cite[\S 7.3.3]{rojas} or \cite[\S 6.5]{DBBNG}.

\subsection{The sources of algorithmic complexity}

The algorithmic complexity of the backpropagation algorithm as presented in \eqref{eqn_error} has
essentially two layers:
\begin{enumerate}[(i)]
\item The complexity of calculating the matrix product.
\item The complexity of the arithmetic operations involved in each step of the matrix multiplications.
\end{enumerate}
As we mentioned in the introduction, one can design smaller and more robust neural networks
performing the same task, but we would need different pruning techniques.  For various network
pruning techniques, see for example \cite{hmd,lkd,tf,ycs}.  The ordinary multiplication algorithm
of matrices of size $n \times m$ and $m \times p$ has the algorithmic complexity of $O(n m p)$.
Any simplification in the structure of the network is going to decrease the algorithmic complexity
as the sizes of the resulting matrices will decrease. Even though there are matrix multiplication
algorithms with better asymptotic complexities (see for instance \cite{legall}), they are generally
impractical due to the large constant factors in their running times. 
It is also not very clear whether these improved matrix multiplications algorithms are well-suited
to the GPU's.

In this note, we propose to focus on the more subtle source of the complexity (ii): the arithmetic
operations.  In algebraic terms, we propose replacing the base field of real numbers and
corresponding arithmetic operations with the \emph{tropical semiring of reals} with its own simpler
arithmetic operations.  Since we only swap the 
field of real numbers with the tropical  ring of real
numbers, we note that any reduction on the algorithmic complexity of arithmetic operations would
not require any structural changes in the backpropagation algorithm. We discuss
these aspects in \S \ref{sec_tr_back}.  Thus, such a swap requires
minimal adaptation in the existing code bases, i.e., swapping the classical linear package with a
tropical linear algebra package.

\begin{remark}
Approximation of  activation function is also notable  source of computational 
complexity as it plays a role in \eqref{eqn_error} as   entries of the matrices $\mathbf{D}^{(k)}$. 
Most of the activation functions are intentionally non-linear, and their evaluations require 
certain precision.  However, tropicalisation of activation functions and especially their approximations
are computationally  less complex as they are realised in terms of piecewise linear functions, see \S 
\ref{sec_corner} below.
\end{remark}

\subsection{Complexities of $\min$/$\max$, addition and multiplication} 
\label{sec_operations}

There are substantial differences in (time) complexities of different the elementary binary
operations. The amount of the  operations to sum two $n$-bits numbers via schoolbook 
addition algorithm is $O(n)$.  Similarly,  the upper bound of the number operations to decide the 
maximum (or the minimum) of the two $n$-bits numbers is similarly $O(n)$. More
importantly,  the average-case complexity of $\max$ (and $\min)$ is  $O(1)$.

By contrast, the schoolbook multiplication for the same size is $O(n^2)$.  While there are other multiplication
algorithms with better algorithmic complexity, such as the Karatsuba algorithm of order
$O(n^{\log_2(3)})$ \cite{karat}, they all have the complexity of $O(n^{\lambda})$ with $\lambda>1$.
It is important to note that these lower algorithmic complexities are achieved asymptotically in
most cases, and they are usually effectively implement for integers.  Moreover, one may argue that
the Kolmogorov complexities of these \emph{asymptotically better} algorithms are not better as they
are significantly difficult to implement.

\subsection{Execution time and energy consumption} 
\label{sec_others}

The actual execution time of a specific code depends on numerous factors such as the processor
speed, the instruction set, disk speed, and the compiler used.  An old rule of thumb in designing
numerical experiments that dictates avoiding multiplications and divisions in simulations in favor
of additions and subtractions in order to improve the actual execution time may heuristically seem
redundant on modern processors  as they closed the time-cost-gap between addition and multiplication
drastically.
However, one can still say the following on the
algorithmic complexity of arithmetic operations:
\begin{itemize}
\item[-] Integer sums and $max/min$ take the same amount of time; 
\item[-] Floating-point operations are   slower than integer operations of the same size;
\item[-] Floating-point multiplications are   slower than the sums. \footnote{See for instance,
    \url{http://nicolas.limare.net/pro/notes/2014/12/12_arit_speed/} and
    \url{https://lemire.me/blog/2010/07/19/is-multiplication-slower-than-addition/} }
\end{itemize}
The energy consumption does not favour the multiplications over the summations either: the required energy
for multiplication is always considerably higher than the summation~\cite[pg. 32]{Horow}.   Even if we
disregard the complexities in CPU designs, both execution time and energy consumption criteria suggest
that the reduction of algorithmic complexity by eliminating the multiplications as much as possible
provide considerable benefits.

\section{Tropical arithmetics and tropical neural networks }

In this section, we introduce the basic notions in tropical arithmetics and tropical geometry, 
then use them to   introduce the tropical neural networks as universal approximators. 


\subsection{Tropical semiring: Arithmetic without multiplication}

The \emph{ tropical semiring} is the limit $\hbar \to +\infty$ of the family
$S_\hbar := (\mathbb{T}, \oplus_\hbar, \otimes_\hbar)$ where $\mathbb{T} := \mathbb{R} \cup \{- \infty\}$
with the following arithmetic operations; for $a, b \in \mathbb{T}$,
\begin{align} \label{eqn_tropical}
  a \oplus_\hbar b  
  & :=
    \begin{cases}
      \log_{\hbar} (\hbar^{a} + \hbar^{b}) & \text{ when } \hbar \in (e,+\infty) \\
      \max\{a,b\}			&  \text{ when } \hbar \to +\infty                        
    \end{cases}\\
  a \otimes_\hbar b &:= \log_{\hbar} (\hbar^{a} \cdot \hbar^{b}) =  a + b. 
\end{align}
%
$S_\hbar$ form a semiring (a ring without additive inverses) and admit the semiring isomorphism
$$
D_\hbar:   (\mathbb{T}, \times, +) \to  S_\hbar 
$$
for any finite value of $\hbar$,  see \cite{ims, viro1}.\footnote{The parameter $\hbar$ is not
just the reminiscent of Planck constant. The tropical limit $\hbar \to \infty$ is essentially  the 
quasi-classical  (i.e., zero temperature) limit of a certain model in quantum mechanics.}

The family $S_\hbar$ in \eqref{eqn_tropical} relates the ordinary addition and multiplication operations
on the set of real numbers with the tropical arithmetic in the limit.  This limiting process is also
called the \emph{Maslov dequantization}~\cite{ims, viro1}.  The tropical limit $S_\infty$ admits a
tropical division, 
however, the subtraction is impossible due to the idempotency of $\oplus_\hbar$, i.e.
$x \oplus_{\infty} x = \max\{x,x\} = x$.  The role of the additive zero is played by $-\infty$, and the
multiplicative unit becomes $0$.

\subsection{The corner locus} \label{sec_corner}

A \emph{ tropical polynomial } is a tropical sum of tropical monomials, therefore, a polynomial of the form
$$
P(\mathbf{x}) = \sum_{(j_1,\dots,j_n) \in V} a_{j_1,\dots,j_n} x_1^{j_1} \cdots x_n^{j_n}, \ \text{with} \  \mathbf{x} := (x_1,\ldots,x_n),
$$
is evaluated in the following form in the tropical  ring
\begin{eqnarray} \label{eqn_variety}
 P^{tr} (\mathbf{x}) &: =&  D_\infty (P(\mathbf{x}))  = 
 \bigoplus_{V} \left( a_{j_1,\dots,j_n} \otimes_{\infty } x_1^{j_1}  \otimes_{\infty }  \cdots  \otimes_{\infty }  x_n^{j_n}   \right)  
\nonumber \\
& =&    \max_{(j_1,\dots,j_n) \in V}  \{ a_{j_1,\dots,j_n}  + \sum j_k x_k  \} 
\end{eqnarray}
where $V \subset \mathbb{Z}^n$ is a finite set of points with non-negative coordinates and the
coefficients $a$'s are tropical numbers.   

The \emph{zero set of a tropical polynomial} $P^{tr}  $ is  the set of  tropical vectors $\mathbf{x} $   
for which either $P^{tr} (\mathbf{x}) = -\infty$, or there exists a pair $i \ne j$ in $V$ such that 
$$    
a_{i_1,\dots,i_n} \otimes_{\infty } x_1^{i_1} \otimes_{\infty } \cdots \otimes_{\infty } x_n^{i_n}   =       
a_{j_1,\dots,j_n} \otimes_{\infty } x_1^{j_1} \otimes_{\infty } \cdots  \otimes_{\infty } x_n^{j_n}. 
$$
Therefore, one can picture the tropical zero set  as the union of 
intersections of hyperplanes defined by the tropical monomials. 
In other words, the tropical zero set  defined by such a polynomial  is given by
the corner locus, that is where the tropical polynomial \eqref{eqn_variety}
is not locally affine-linear.


\subsection{ Rectified linear unit   and its tropical degeneration}
In order to describe the tropical degeneration of the rectified linear unit (ReLU)
\begin{equation} \label{eqn_relu}
\sigma ( \mathbf{x}  ) 
= \max\left\{0, b+ \sum_{i=1}^n a_i x_i \right\}
\text{ where }
\mathbf{x} = (x_1,\dots,x_n) \in \mathbb{R}^n, 
\end{equation}
we consider the $\log$-$\log$ plot of this activation function as family with respect to a parameter
$\hbar \in [e,\infty)$. The transition to the log paper corresponds to the change of coordinates: 
$$
v_\hbar = \log_\hbar (y), \ \ \text{and} \ \  u_{i} = \log_\hbar (x_i)
$$
  Then, we simply obtain
$$
v_\hbar = \log_{\hbar} ( \sum_{i=1}^n  \hbar^{\alpha_i} \hbar^{u_i} +\hbar^\beta )
$$ 
where  $b = \hbar^{\beta}$ and $a_i = \hbar^{\alpha_i}, i= 1,\dots,n$.  In its
domain, the tropical limit $\hbar \to \infty$  of this $\log$-$\log$ graph of the function
ReLU becomes
\begin{equation} \label{eqn_tropical_unit}
\sigma^{tr} := \lim_{\hbar \to \infty} v_\hbar  = \max\{b, \max_{i=1,\dots,k}\{a_i + x_i\} \}.
\end{equation}
In other words, the tropical degeneration of ReLU is another ReLU in an appropriately defined domain.

\subsection{Multi-layered neural networks as universal approximators }
Multi-layered neural networks are used for approximating an unknown function described by a sample
of points in a large affine space $\mathbb{R}^n$ called a \emph{data set}.  The theoretical
underpinnings of such an approximation goes as far back as Kolmogorov~\cite{spe,Cy,Hor}.  We know
that for a given   bounded, and monotonically-increasing continuous function $\sigma$
and a compact domain $\Omega\subseteq\mathbb{R}^n$, any continuous function
$f\colon \Omega\to \mathbb{R}$ can be approximated by finite linear sums of the form
\begin{equation}
 \label{eqn_onelayer}
  F(x) = 
\sum_{i =1}^{N} \beta_i \ \sigma (b_i + \sum_{j=1}^n  a_{ij} x_j ).  
\end{equation}
 As a result, the neural networks with different activation function such as the logistic function, 
 $\arctan$, $\tanh$, SoftPlus etc., are often thought as the universal approximator of continuous 
 functions~\cite{rojas, DBBNG}.

We view neural networks as a concrete computational manifestation of this \emph{Universal 
Approximation Theorem}  where $\sigma$ plays the role of the activation function in a neural 
network.   Our focus in this
 paper is to develop better computational methods in achieving such approximations.

\subsection{Approximation via tropical neural networks}
\label{sec_approx}

The rectified linear unit (ReLU) we defined in \eqref{eqn_relu} also lies in this class of
functions that can be used for approximations of continuous and the piecewise smooth functions on
any compact domain in $\mathbb{R}^n$.  As we observe in \eqref{eqn_tropical_unit} that ReLU can
also be represented by using tropical unit on compact domains.  Thus, we can simply state that the
ReLu and the tropical units $\sigma^{tr}$ in \eqref{eqn_tropical_unit} are equivalent from the
perspective of aproximation theory:

\noindent
\shadowbox{
  \begin{minipage}{\dimexpr\textwidth-\shadowsize-2\fboxrule-2\fboxsep}
    \begin{proposition}\label{prop:star}
     Multilayered feedforward neural networks using tropical unit $\sigma^{tr}$ in
     \eqref{eqn_tropical_unit} as the activation function can give arbitrarily close approximations
     for any continuous and piecewise smooth function on any compact domain in $\mathbb{R}^n$.
    \end{proposition}
\end{minipage}
}

For the deep ReLU networks, these approximations work as effectively.  For details see \cite{peter,ya}.

\section{Backpropagation in tropical setting }

\subsection{Reformulating a classification problem as a topological problem}

Consider a smooth map $ f: \mathbb{R}_+^n \to \mathbb{R}$ defined over the cone
$\mathbb{R}^n_+ = (0,\infty)^n$, and consider the zero locus $Z_f := \{x \in \mathbb{R}^n \mid f(x) = 0\}$
of $f$.  Let us define a map
$$ F: \mathbb{R}_+^n \setminus  Z_f \to H_0( \mathbb{R}_+^n \setminus Z_f) $$
from the complement of $Z_f$ to the set of connected components of $ \mathbb{R}_+^n \setminus Z_f$. The
map $F$ simply sends each elements $x \in \mathbb{R}_+^n \setminus Z_f$ to their homology classes
$[x] \in H_0( \mathbb{R}_+^n \setminus Z_f)$, i.e. the connected component of $ \mathbb{R}_+^n \setminus Z_f$ 
containing $x$. For basics 
of  (co)homology theory, see \cite{bott}. 

Now, given an arbitrary finite data set  (or a finite set of compact subsets) $\Omega$ in $\mathbb{R}_+^n$,
we can reformulate a classification problem
\begin{equation} \label{eqn_classify}
N: \Omega  \to \{1,\cdots k\}
\end{equation} 
as a problem of finding a smooth map $f$ satisfying the following property: for $x,y \in \Omega$,
$$ N(x) = N(y) \iff F(x) = F(y). $$
Clearly, this reformulation requires that $ \mathbb{R}_+^n \setminus Z_f$ has at least $k$ connected
components so that $F$ can realise the same classification problem \eqref{eqn_classify}.

\subsection{Backpropagation and its tropicalisation}
\label{sec_tr_back}

Let $f$ be a function which realises the classification problem in \eqref{eqn_classify}, and let
$\{f_i \mid i =0, \dots , \infty\}$ be a sequence of deep ReLU networks coming from the
backpropagation algorithm in \eqref{eqn_error}
$$
\lim_{k \to \infty} f_k = f.
$$
Each function $f_k$ is piecewise linear and continuous as they are realised by multilayered ReLU
networks. See Proposition~\ref{prop:star}.

The corner locus of each $f_k$ is mapped into the corner locus of the tropical limit
$f^{tr}_k = \lim_{\hbar \to \infty} f_k$. In addition to the tropical images of these existing
corner locus, the tropical limit of $f_k$ form additional corner locus which is the tropical limit
of the zero locus $Z_{f_k} = \{f_k = 0\} \subset \mathbb{R}^n_+$ under tropical degeneration. This
new corner locus $Z^{tr}_{f_k}$ is the tropical zero set of $f^{tr}_k$ defined by
$\lim_{\hbar \to \infty} f_k$.

\begin{lemma}
  The  zero locus $Z_{f_k}$ of $f_k$ is homeomorphic to the   tropical set $Z^{tr}_{f_k}$. 
\end{lemma}

This statement follows from the fact that the map
\begin{equation} \label{eqn_log} \text{Log}_\hbar: \mathbb{R}^n_+ \to \mathbb{R}^n: (x_1,\dots,x_n)
  \mapsto (\log_\hbar (x_1),\dots,\log_\hbar (x_n))
\end{equation}
is a diffeomorphism for all $\hbar < \infty$.  Then the image $\text{Log}_\hbar (Z_{f_k})$ is homeomorphic
to $Z_{f_k}$ for any finite $\hbar$.  For the tropical limit $\hbar \to \infty$, we use the fact
$$ \lim_{\hbar \to \infty} \frac{d}{d \hbar} \text{Log}_\hbar = 0 $$
to show that $Z^{tr}_{f_k}$ is homeomorphic to $\text{Log}_\hbar (Z_{f_k})$ for sufficiently large
$\hbar$.

We note that this statement is in fact a special case of Viro's theorem \cite{viro}. Viro observed
that tropical degeneration preserves the topology of real algebraic varieties.  He developed a
method, known as \emph{Viro patchworking}, that combinatorially constructs any real algebraic
variety with a prescribed topology.  We believe that Viro's method in its general form can also be
directly used in classification problems in \eqref{eqn_classify}.
 
\begin{corollary}\label{cor:TropicClassification}
  The classification $N: \Omega \to \{1,\cdots k\}$ in \eqref{eqn_classify} can be realised by the
  tropical set
  $$  f^{tr} = \lim_{k \to \infty} f^{tr}_k $$ 
  as the tropical set that $ f^{tr} $ defines is homeomorphic to $Z_f$.
\end{corollary}
 
Corollary~\ref{cor:TropicClassification} essentially means that we can simply apply
$\text{Log}_\hbar$ \eqref{eqn_log} to each step of the backpropagation algorithm given in
\eqref{eqn_error} to define the tropical version of the backpropagation algorithm.  This is done by
taking tropical image of each entry of the matrices, and then replacing the matrix addition and
multiplication by their respective tropical arithmetic operations $\oplus_\infty$ and
$\otimes_\infty$ in \eqref{eqn_tropical}: Let $\mathbf{A} = [a_{ij}]$  and $\mathbf{B} = [B_{ij}]$
be $n \times m$ matrices. The \emph{ tropical matrix sum}, $\mathbf{A} \oplus_\infty \mathbf{B}$,  is then 
obtained by evaluating the tropical sum of the corresponding entries,
$$
(\mathbf{A} \oplus_{\infty} \mathbf{B})_{ij} := a_{ij} \oplus_\infty b_{ij} = \max \{a_{ij}, b_{ij} \}. 
$$
The \emph{ tropical multiplication $\mathbf{A} \otimes_{\infty} \mathbf{B}$ of two  matrices} $\mathbf{A} = [a_{ij}] \in \mathbb{R}^{m \times n}$
and $\mathbf{B} = [b_{ij}] \in \mathbb{R}^{n \times p}$ is the given by the matrix 
$\mathbf{C} = [c_{ij}] \in \mathbb{R}^{m \times p}$ with entries
$$
c_{ij} := \oplus_\infty (a_{ik} \otimes_\infty b_{kj}) = \max_k \{a_{ik}  + b_{kj} \}. 
$$

By using the error term \eqref{eqn_error} and tropical linear algebra defined above,  
we formulate tropical gradient descent  iteration  as follows
\begin{eqnarray} \label{eqn_iteration}
\mathbf{W}^{(k)}_{\text{new}}  & =  & \mathbf{W}^{(k)} \oplus_\infty \Delta  \mathbf{W}^{(k)}  \nonumber\\
						& = &  \mathbf{W}^{(k)} \oplus_\infty - \epsilon   \left(
						\mathbf{D}^{(k)}  \otimes_{\infty}    \mathbf{W}^{(k+1)}  
						 \otimes_{\infty}   \cdots   \otimes_{\infty}   \mathbf{D}^{(l)}   \otimes_{\infty}  \mathbf{W}^{(l+1)} \otimes_{\infty}   \mathbf{e} \right). 
\end{eqnarray}

\noindent 
\shadowbox{
  \begin{minipage}{\dimexpr\textwidth-\shadowsize-2\fboxrule-2\fboxsep}
    \begin{corollary}
      The tropical backpropagation algorithm in \eqref{eqn_iteration} has a lower algorithmic complexity relative to
      the vanilla backpropagation algorithm.
    \end{corollary}
  \end{minipage}
} 

This statement   follows from the construction of tropical backpropagation which 
simply eliminates the complexities of multiplications.

\begin{remark}
  On a machine with a fixed register size (the number of bits available to represent a number),
  the improvements we get by swapping the ring of reals with the tropical semiring of reals will
  only effect the \emph{the constant factor} of the complexity of matrix operations.  On a
  classical machine with 128-bit registers, the complexity of ordinary multiplication of two
  $n \times n$ matrices is $O(n^3)$ with constant factor of $2^{14}$ while the same multiplication
  on a tropical  machine with the same size registers will take again $O(n^3)$ time but with the
  constant factor of $2^7$.
\end{remark}

\begin{remark}
  The tropical linear algebra is efficient and lowers the algorithmic complexity of large matrix
  operations, and therefore, fits well with the backpropagation and may be used in other
  applications.  However, there are also serious limitations as it does not admit matrix inversion
  due to its idempotent nature.
\end{remark}

\section{Conclusion}

In this note, we defined the tropical limit of the ReLU function which is used in many neural
network models.  We also showed that the multilayered feedforward neural networks using this
tropical unit carry the properties of the universal approximator.  With further analysis, we
established that the topology of the zero loci of functions realised by the multilayered neural
networks do not change when their tropical limit is considered.  This observation allowed us to
tropicalize the backpropagation algorithm solving any classification problem. The tropical
backpropagation algorithm is simply obtained from the classical backpropagation algorithm in
\eqref{eqn_error} and replacing the matrix addition and multiplication with their tropical
analogues based on the tropical arithmetic operations $\oplus_\infty$ and $\otimes_\infty$ in
\eqref{eqn_tropical}.

As the tropical backpropagation algorithm works over the tropical semiring, it comes with a
considerably algorithmic advantages with almost no drawback:
\begin{itemize}
\item Tropicalization reduces the algorithmic complexity significantly by eliminating
  multiplications.
\item The performance gain out of tropicalization does not come with a cost of substantial change
  in the existing code, since it only requires a swaping an ordinary linear algebra library with an
  appropriate tropical linear algebra library.
\end{itemize}

\subsubsection*{Acknowledgments}
I  wish to thank to Grisha Mikhalkin, not only for introducing me to the tropical geometry over the 
years, but also  for inviting me over to Geneva (in various occasions) and listening
my half baked ideas with Ilia Itenberg patiently. I also thank to Yi\u{g}it G\"und\"u\c{c} who
introduced me to the concepts of computational complexity decades earlier. 

I am grateful to Atabey Kaygun, for being an unwearying friend and collaborator. Main idea of
this note appeared first during a discussion with Atabey, and he read and commented on
all versions of it, i.e., he is a secret author of this paper. Saying that, all mistakes are mine, mine only.



\end{document}